\documentclass[a4paper,11pt]{article}
\pdfoutput=1 

\usepackage{jinstpub} 

\usepackage{lineno}
\nolinenumbers

\title{A proton-recoil track imaging system for fast neutrons: the RIPTIDE detector}

\author[a]{P. Console Camprini,}
\author[b,c]{F. Leone,}
\author[d,e]{C. Massimi,}
\author[b,c]{A. Musumarra,}
\author[b,c]{M.G.~Pellegriti,}
\author[e]{C.~Pisanti,}
\author[b,c]{F. Romano,}
\author[d,e]{R. Spighi,}
\author[a]{N. Terranova}
\author[d,e,1]{and M. Villa\note{corresponding author}}

\affiliation[a]{Agenzia nazionale per le nuove tecnologie, l’energia e lo sviluppo economico sostenibile, Frascati, Italy}
\affiliation[b]{Istituto Nazionale di Fisica Nucleare, Sezione di Catania, Catania, Italy}
\affiliation[c]{Dipartimento di Fisica e Astronomia, Università di Catania, Catania, Italy}
\affiliation[d]{Istituto Nazionale di Fisica Nucleare, Sezione di Bologna, Bologna, Italy}
\affiliation[e]{Physics and Astronomy department, Università di Bologna, Bologna, Italy}

\emailAdd{mauro.villa@bo.infn.it}

\abstract{Fast neutron detection is often based on the neutron-proton elastic scattering reaction: the ionization caused by recoil protons in a hydrogenous material constitutes the basic information for the design and development of a class of neutron detectors. Although experimental techniques have continuously improved, proton-recoil track imaging remains still at the frontier of n-detection systems, due to the high photon sensitivity required. Several state-of-the-art approaches for neutron tracking by using n-p single and double scattering – referred to as Recoil Proton Track Imaging (RPTI) – can be found in the literature. So far, they have showed limits in terms of detection efficiency, complexity, cost, and implementation. In order to address some of these deficiencies, we have proposed RIPTIDE a novel recoil-proton track imaging detector in which the light output produced by a fast scintillator is used to perform a complete reconstruction in space and time of the interaction events. The proposed idea is viable thanks to the dramatic advances in low noise and single photon counting achieved in the last decade by new scientific CMOS cameras as well as pixel sensors, like Timepix or MIMOSIS.
In this contribution, we report the advances on the RIPTIDE concept: Geant4 Monte Carlo simulations, light collection tests as well as state-of-the-art approach to image readout, processing and fast analysis.}


\begin{document}
\maketitle
\flushbottom

\section{Introduction}

In the field of neutron detection, many different techniques have been developed over the years to measure the yield of neutrons and possibly their energies. Interaction probability with matter and detection efficiencies are a strong function of the energy and usually, for fast neutrons, the detection efficiency is low. Since neutrons are neutral particles, there is not the possibility to follow their track and the measurement of the neutron impinging direction is usually not possible with a single detector, but relies on the knowledge of the neutron production point. Still there are several interesting cases where the detection of the neutron track direction is crucial in order to distinguish interesting neutrons from background signals. This is the case, for example, for solar neutrons with energy $E>1$~MeV: the only known measurement~\cite{law} done so far, from the MESSANGER probe, provided no conclusive information since it was not possible to separate neutrons coming from the sun from neutrons produced in secondary interactions on the probe material~\cite{ger}. Nowadays this field of detector research is lively advancing due to different proposals, all based on recoil proton scattering. These new attempts are based on scintillating detectors or gaseous detectors enriched in hydrogen where a fast neutron can have a $n$-$p$ interaction, therefore it can produce a moving charge in the active detector volume. The proposed detectors~\cite{hu, mondo1, mondo2, sontrac} suffer in general from low efficiencies or the capability to determine only a single $n$-$p$ interaction, thus limiting strongly their tracking capability. Here we present a novel design~\cite{mus,cm} of a neutron tracking detector, called RIPTIDE (RecoIl Proton Track Imaging DEtector) that overcomes both limitations.

\section{Proton recoil techniques and neutron tracking}

Fast neutrons impinging on hydrogen rich materials like organic scintillators or organic materials can give rise to $n$-$p$ scattering events. For kinetic energies up to the opening of inelastic channels (280 MeV), the $n$-$p$ elastic cross section is dominating over or of the sime size of other elastic and inelastic cross sections on Carbon or Oxygen nuclei that might be present in the medium. The scattering is therefore a simple two body process with particles with almost equal masses. The energy transfer to the target proton is described simply as $E_{\rm p} = E_{\rm n} \cos^2 \theta$, where $E_{\rm n}$ is the impinging neutron kinetic energy, $E_{\rm p}$ is the recoil proton kinetic energy and $\theta$ is the scattering angle. Now lets suppose that a given neutron undergoes two elastic scatterings and that it is possible to measure the energies, the starting points and the directions of the two recoil protons. In these conditions it is possible to reconstruct the flight path of the neutron between the two interactions, to measure the scattering angle of the second proton and measuring its energy, it is possible to determine the neutron energy after the first scattering. By measuring direction and energy also for the first recoil proton, the final state of the first interaction is fully known. By energy and momentum conservation it is possible then to determine the energy and the direction of the impinging neutron. 

\begin{figure}[htbp]
\centering
\includegraphics[width=.9\textwidth,clip]{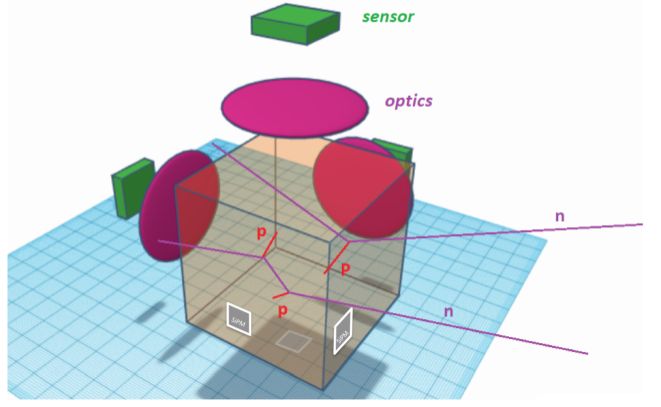}
\caption{\label{fig:riptide} Minimal configuration of the RIPTIDE detector with a cube scintillator, SiPMs, lenses and optical pixel sensors. As examples, on the drawing it is possible to see a neutron interacting once and another interacting twice in the scintillator, producing recoil protons.}
\end{figure}

\section{Detector basic principles}

In our application we are considering for the active volume a cube of plastic scintillator with a H:C ratio of 1.1 and a typical length of $L=6$~cm. Detectable fast neutron interactions in this volume are those on Hydrogen and, for kinetic energies up to 400 MeV, they are mainly of elastic scatterings. Recoil protons move inside the cube releasing energy in the form of scintillating light. For long enough ranges, the light production is maximal at the end of the range (Bragg peak) and it is possible to define a starting and an ending track point. The track length provides the proton energy measurement; the beginning of the track (defined as the low intensity part of the track) defines the neutron interaction point and the proton initial direction. In order to measure these characteristics, we need to reconstruct how the light is produced along the track, therefore we need to observe an image of the proton track. For this reason the scintillator cube should be seen by cameras able to record the produced light and to have an actual image of the track. In order to be able to collect as much light as possible an optical system consisting of at least a lens is coupled to an imaging light detection device. Two or better three of such light detection devices are needed in order to perform the 3D reconstruction of the proton track. The scintillation cube will also be equipped with non imaging sensors (such as commercial SiPMs) to have a fast signal connected to the neutron scattering inside the cube. The whole detector is sketched in Fig.\ref{fig:riptide} where, as an example, the interactions of two neutrons having a single and a double scattering are also shown.

\begin{figure}[htbp]
\centering
\includegraphics[width=.6\textwidth,clip]{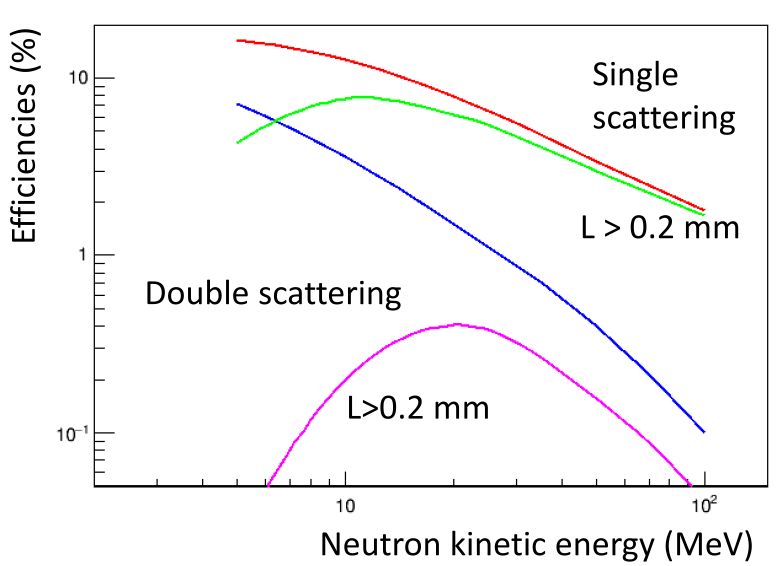}
\caption{\label{fig:eff} Probability of interaction and detection efficiency for neutrons as a function of the kinetic energy. The red line refers to single scattering into the cube; the green line refers to recoil proton ranges greater than 0.2 mm (minimum detectable). The blue line refers to double neutron scattering into the cube, while the purple line is obtained requiring that both recoil protons have a range greater than 0.2 mm.}
\end{figure}

Several aspects have to be accounted for in order to have a reliable and working detector. For a rough estimation of the minimal performance we consider the detection of protons with path length in the scintillator longer than 0.2 mm. This is considered the minimum length for which we'll be able to identify a track direction. This defines the minimal detectable proton energy that is 3.5 MeV. Since in typical organic scintillator the light in the cube can be evaluated as $3\cdot 10^3$ photons per MeV of deposited energy, we'll have typically at least $10^4$ photons over the entire solid angle. This light will be collected outside the cube by optical systems (lenses): necessarily we need not to rely on internal light reflection, that will be considered as a source of noise. The cube will be therefore covered by light absorbing material except for three opening (windows) from which we'll see the cube interior and for the places where SiPMs will be placed to detect direct scintillation light.  Since SiPMs can be sensitive to single photons and at least 3 SiPMs will be used, there will be enough signal on each one to use their signal for triggering.
In fig. \ref{fig:eff} the probability of interaction for single and for double proton scattering are shown together with the probability of having recoil proton tracks longer than 0.2 mm.

For what regards the optical system, the exact position and characteristics of the lenses will be defined as the optimal compromise between the need to collect as many photons as possible 
(a large lens, placed close to the cube) and a deep field of view (long focal length). A typical lens distance is in the range 5-8 cm from the cube center, which will reduce the solid angle coverage by a factor of at least 20.
The collected light could then be put in input to an image intensifier, made by a microchannel plate or put directly to an high sensitivity imaging sensor.

\section{Sensor chips}

\begin{figure}[htbp]
\centering
\includegraphics[width=.7\textwidth,clip]{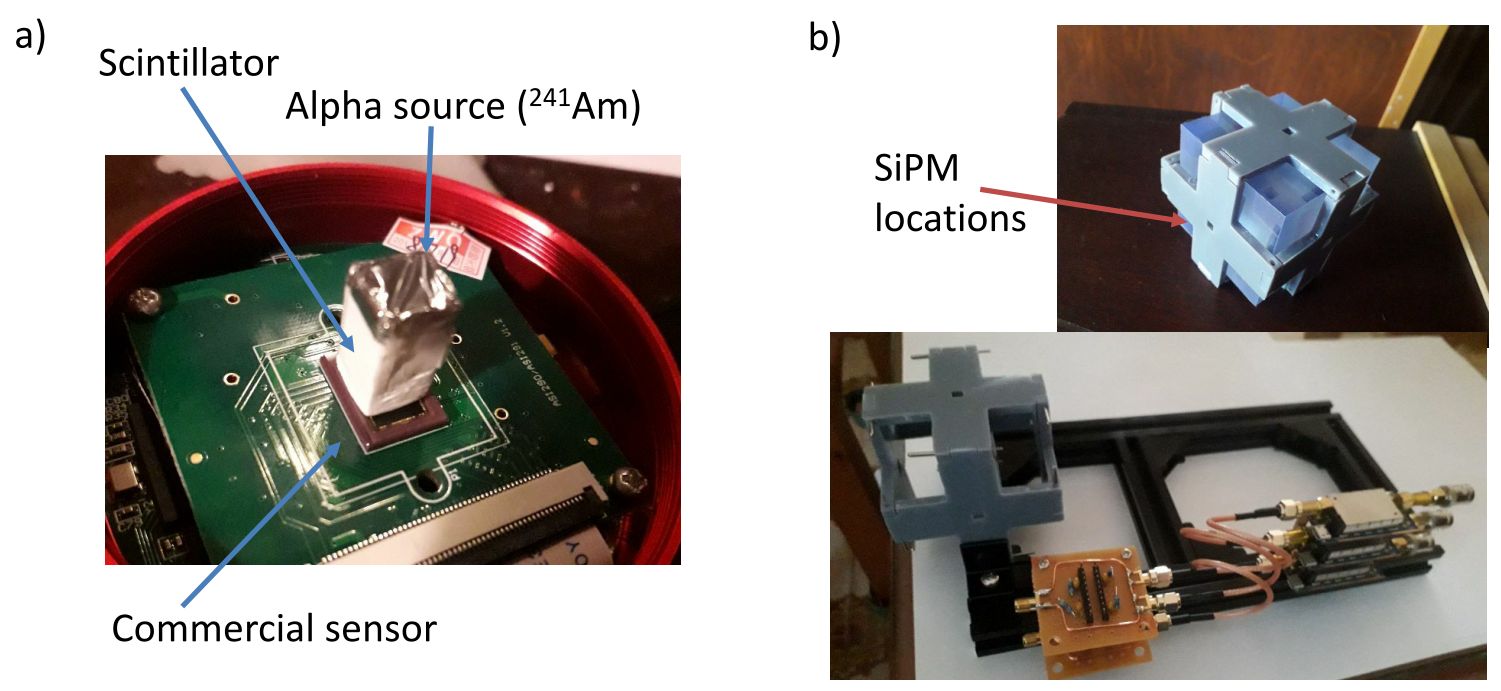}
\caption{\label{fig:cube} a) A test with a scintillator, a commercial CCD sensor and a neutron source; b) Top: the scintillator cube with the plastic holding to host the cube and the SiPM sensors; bottom: the cage and some analog electronics.}
\end{figure}

The choice of the sensor will characterize the time and spatial performances of the whole setup,
so it is the most critical element. The first lab studies have been performed by a small scintillator
coupled to a $^{241}$Am alpha source coupled with a commercial camera designed for astrophotography
and equipped with a IMX290MM Sony CMOS sensor. The arrangement can be seen in Fig.~\ref{fig:cube}.a. This
sensor is characterized by a large high efficiency and low noise (1 e-rms).
The maximum acquisition frame rate is 170 fps, much more than usually needed
for astrophotography, but not enough for our application.
In Fig.~\ref{fig:cube}.b, top part, there is a picture of the scintillator cube (6 cm side length), with a plastic holding designed to host 6 SiPMs for overall signal studies. In Fig.~\ref{fig:cube}.b, lower part, there is a picture of the holder of the cube together with a minimal electronics for SiPM powering and reading out.

There are several problems related to off-the-shelf CCD sensors for our application, that makes then not suitable in general. In a typical organic scintillator the light is produced in a time window that can be of the order of few tens of nanosecond, up to 100 ns. Any light integration on a longer time period will increase the noise in each pixel. For this reason we've searched for imaging pixel sensors having high frame rates (up to few kfps) in order to be able to reduce at minimum the random telegraph noise in each pixel and to be able to acquire more neutron interaction events in each second. Another design requirement is related to the dead time between different frames, that will constitute a source of inefficiency. This is a characteristics that is intrinsic to the readout architecture of the sensor itself. At the moment we are considering two different families of sensors: a) commercial high rate CCD/CMOS sensors and b) specific particle detector pixel sensors.

Commercial high rate CCD/CMOS sensors are designed mainly for line production checks or for automotive. They are employed usually in high luminosity environments. Notwithstanding this in order to collect the maximum light in each frame two interesting characteristics for our application are usually present: the sensor is illuminated from the back for a maximum light collection and, part of the area is used to store the charge collected from the previous exposure. This allows the parallelisation of the two phases of exposure and reading, reducing the dead time at minimum. A layer of microlenses helps in directing the light on the active area, enhancing the overall efficiency. Due to the sparse photon countings no color filter should be used for our application. Sensors matching all these requirements are for example the LUPA3000 and the LU19HS sensors developed by OnSemi~\cite{semi} that can reach up to 2kfps at full resolution at about 2 Mpixels. 
In order to improve the photon statistics it is possible to use an image intensifier between the lens and the camera. It is well known that Micro Channel Plates (MCP) provide suitable spatial resolution (about 10-100 $\mu$m) and a large detection surface. Timing is a key feature of such a device, reaching time resolution of 100-200 ps. They have been extensively used in
nuclear physics, neutron, UV and X-ray imaging applications since decades \cite{pie,mus2,sieg}. MCP implementation in RIPTIDE fulfills the basic requirement of event-by-event n-tracking with sub nanosecond time resolution. Concerning light detection efficiency, the use of several photocathodes can be investigated, considering that state-of-the-art photocathodes for UV astronomy can achieve QE in the range 20-60\% (GaN, Cs, Diamond) \cite{mus2}.
With this approach a simple MCP or a chevron-configuration MCPs can provide enhanced images on a green phosphorous screen. Images on the screen can then be recorded by a high frame rate camera.

A complementary approach, currently under consideration, will use the image intensifier made by microchannel plates, but instead of placing a phosphorous screen to receive the electrons from the MCP, a particle pixel sensor could be used, has it has been done for \cite{timepix}. For this configuration there are several interesting chips to consider. For example, the latest version of the Timepix chip \cite{tp4} is a large area (7.4 cm$^2$) 229k pixel chip with subnanosecond timing capability and an highly configurable read out. The Ultimate M28 chip \cite{m28}, of the Mimosa series, is another ready solution: it has about 1 Mpixels and it provides a simple hit/not hit information on each pixel. The pixel matrix is read out continuously, having a rolling shutter architecture, in a fixed reading time of 185 $\mu$s. It has therefore a maximum full resolution frame rate of 5.4 kfps. A new sensor, called MIMOSIS \cite{mim}, similar to M28 for the architecture, but better for time performances is currently under development for the Compressed Barionic Experiment \cite{CBM}. This is a 5.4 cm$^2$ sensor with 0.7 Mpixels and a timestamping capability down to 5 $\mu$s, allowing a remarkable 200 kfps, with a sparsified readout.

\begin{figure}[htbp]
\centering
\includegraphics[width=.7\textwidth,clip]{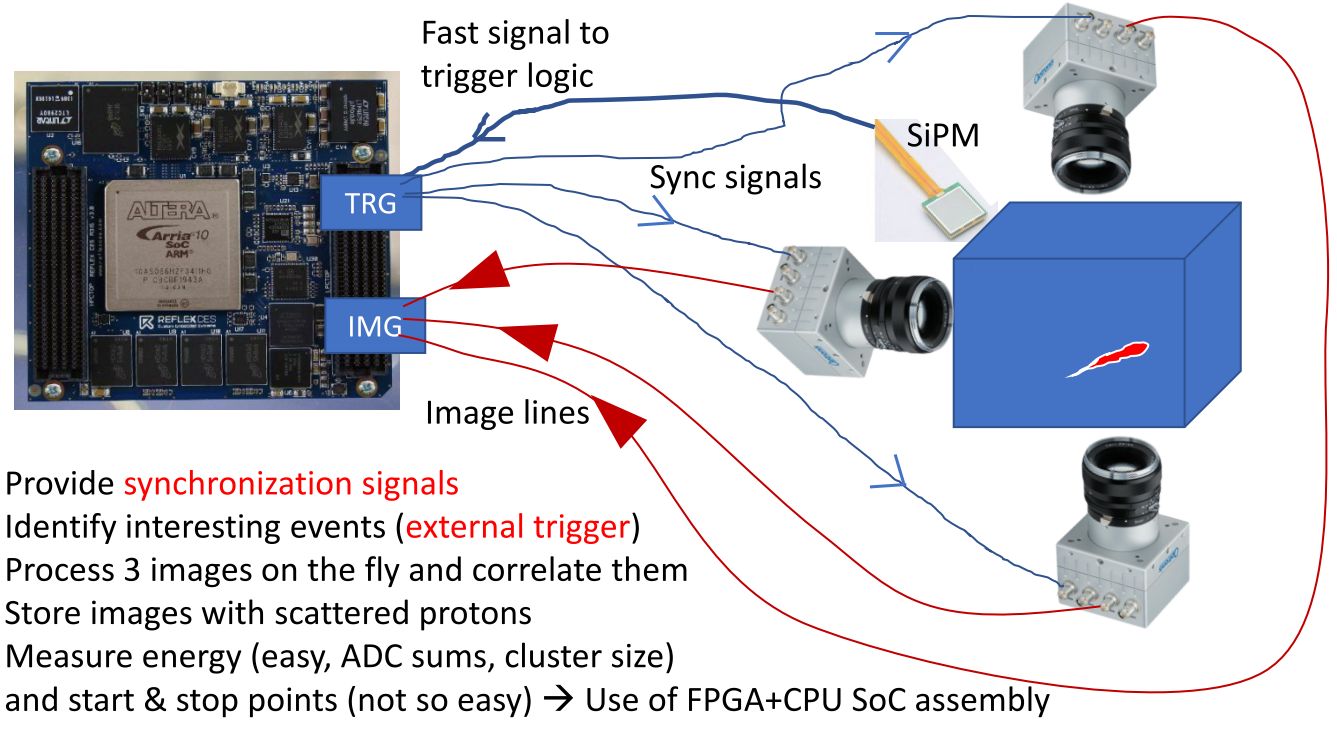}
\caption{\label{fig:ele} Schematics of the electronic connections: The scintillator cube is seen by three cameras and at least one SiPM. Cameras can be synchronized. SiPM signal provides a trigger to tag interesting frames received from the acquisition board.}
\end{figure}

\section{Electronics}

The RIPTIDE detector is completed by its electronics simply sketched in fig. \ref{fig:ele}. At the core of the electronics is placed a general purpose highly programmable board able to sustain high input data rates. The board will be connected to the SiPMs and to the cameras. It will receive digitized signals from the SiPMs that will be used to tag interesting frames. If possible the cameras will receive a synchronization signal or a common clock to have the starting of the frames synchonized over the different cameras. The cameras will send continuously the recorded frames to the board. The input bandwidth for each camera/sensor can be very different depending on the sparsification capabilities of the sensor itself. For a typical image sensor with 1 Mpixels at 1 kfps, at least 1 GB/s input rate is needed. For sparsified readout architectures like those of M28 or Timepix4 input rates of 10 MB/s are usually enough in our application. Since we expect that most of the frames will be empty, an immediate processing of the recorded images is mandatory. For this reason, an optimal choice for the electronic board is a last generation of a System-on-a-Module (SOM) board with a FPGA and an embedded ARM CPU.  Simple and fast noise subtraction level and feature extraction algorithms will be applied in the FPGA to each acquired image, in order to disentangle empty and event frames. Moreover, events with at least two sensors having non-empty frames and/or with a fast SiPM signal in coincidence will be sent to the
CPU for further processing, selection and data delivery to the final storage. The 1 Gbps ethernet ports in the SOM will be used for this purpose. Typical SOM boards based on Altera Arria 10 or Xilinx Kintex 7 FPGA are well suitable for our purposes. A firmware/software code providing the basic treatment of the input data is currently being developed. From the FPGA side, it collects frames from up to 4 independent channels, it assign time-tags to them and it stores on a temporary RAM sorted by their time-tag. The RAM is handled like a circular buffer and continuously filled from one side and emptied from the other. By external SiPM signals or by feature extraction on each frame (total light collected) an internal trigger is issued and a range of interesting frames related to that trigger is identified. This information is available to the CPU side of the SOM, which is in the condition to read the shared RAM and extract the triggered frames and send them to a storage system via a 1 Gbps ethernet connection.


\section{Conclusions}

A new detector providing tracking capabilities for fast neutrons using in an original way high frame rate imaging sensors has been described. The RIPTIDE detector is based on a 6 cm side cube of scintillating material that has a maximum efficiency for double neutron scattering at 20 MeV of neutron kinetic energy. The different parts of the detector have been discussed. Few options are still open expecially on the most critical part, the imaging sensors, that have to fullfill demanding requirements: high frame rate (>1kfps), large area (>1 Mpixels) and low photon count. There are interesting applications of a detector of this type that range from energy neutron measurements in laboratory up to solar neutron identification and solar neutron spectroscopy in space. 

\acknowledgments
The authors acknowledge the use of laboratory and computational resources from the Open Physics Hub (https://site.unibo.it/openphysicshub/en) at the Physics and Astronomy Department in Bologna.



\end{document}